\documentclass[a4paper, amsfonts, amssymb, amsmath, reprint, showkeys, nofootinbib, twoside]{revtex4-2}
\usepackage[english]{babel}
\usepackage[utf8]{inputenc}
\usepackage[colorinlistoftodos, color=green!40, prependcaption]{todonotes}
\usepackage{amsthm}
\usepackage{mathtools}
\usepackage{physics}
\usepackage{xcolor}
\usepackage{graphicx}
\usepackage[left=23mm,right=13mm,top=35mm,columnsep=15pt]{geometry} 
\usepackage{adjustbox}
\usepackage{placeins}
\usepackage[T1]{fontenc}
\usepackage{lipsum}
\usepackage{csquotes}
\usepackage[pdftex, pdftitle={Article}, pdfauthor={Author}]{hyperref} % For hyperlinks in the PDF
\begin{document}
\title{Mass-conservation increases robustness in stochastic reaction-diffusion models of cell crawling}

\author{Eduardo Moreno}
\author{Sergio Alonso}
    \email[Correspondence email address: ]{s.alonso@upc.edu} % Your name
    \affiliation{Department of Physics, Universitat Politècnica de Catalunya, Barcelona, Spain}

\date{\today} % Leave empty to omit a date

\begin{abstract}
The process of polarization determines the head and tail of single cells. A mechanism of this kind frequently precedes the subsequent cell locomotion and it determines the direction of motion. The process of polarization has frequently been described as a reaction-diffusion mechanism combined with a source of stochastic perturbations. We selected a particular model of amoeboid cell crawling for the motion of \textit{Dictyostelium discoideum} and studied the interplay between pattern formation and locomotion. Next, we integrated the model in a two-dimensional domain considering the shape deformations of the cells in order to characterize the dynamics. We saw that the condition of pattern formation is finely tuned and we propose a modification based on the use of a mass-conservation constraint to substantially increase the robustness of the mathematical model.

\end{abstract}


\maketitle

\section{Introduction}

An intensive use of physical and mathematical theories on pattern formation in extended systems \cite{cross1993pattern,koch1994biological}, gave rise to valuable arguments to explain the formation of certain biological structures. 
Such mathematical mechanisms have permitted the modeling of processes on very different spatial and temporal scales: from the formation of the skin in fishes \cite{kondo2010reaction}, to the definition of the direction in embryonic developing \cite{gross2019guiding}.

Cell migration is a common phenomenon that is present both in prokaryotic and eukaryotic cells. %as well as in some bacteria. 
Living cells migrate to perform different tasks such as food targeting, wound healing and immune response. 
Independently of the presence of an external signal, before moving, cells need to define the direction to follow. To do so, they first define the front and the back of the cell. 
The process of formation of a polar direction inside a single cell is commonly known as cell polarization \cite{mogilner2012cell} and it is a typical example of pattern formation at the cellular level \cite{camley2017physical}. 

Several mathematical models have been developed to explain polarization of single eukaryotic cells   \cite{rappel2017mechanisms,jilkine2011comparison,beta2017intracellular}.
Some models rely on a local excitation, which, combined with global inhibition makes the cell respond to external gradients \cite{xiong2010cells}, while others rely on the accumulation of a certain biochemical components to guide the motion of the single cells \cite{mori2008wave}. Normally the accumulation responsible for this second mechanism is combined with a conservation constraint because the process of polarization is fast in comparison to the production of new biochemical components. 
With this restriction, the mechanism of pattern formation inside living cells together with the constraint of mass conservation is analogous to a process of coarsening of the initial nucleus of components \cite{bergmann2019system} and gives rise to phase separation \cite{brauns2020phase} and to models of pattern formation \cite{halatek2018rethinking}. 
Several simple models of intracellular pattern formation have appeared that include the conservation restriction \cite{mori2008wave,alonso2010phase,trong2014parameter}.

%%- Cell polarization and motion

%%- Models of cell polarization\cite{rappel2017mechanisms} \cite{jilkine2011comparison}

%%- Models of cell crawling 

%%- Dicty cells and models \cite{xiong2010cells} \cite{iglesias2008navigating}

%%- Bistable Models based in patches \cite{hecht2010transient} \cite{van2017coupled}   extension to 2d membrane dynamics together with the global resource limitation in limiting the extensions of the pseudopods. \cite{hecht2011activated}. Similar membrane dynamics based of moving conected points... 
%%\cite{neilson2011chemotaxis}

%- Phase field models to extend into 2D \cite{alonso2018modeling}  \cite{moreno2020modeling}

%-  Mass conservation models ... \cite{brauns2020phase}  \cite{halatek2018rethinking}

%\cite{mori2008wave}, \cite{otsuji2007mass},  

%Coarsegraining and mass conservation \cite{bergmann2019system}

%Several models have been proposed to study and reproduce cell motility \cite{jilkine2011comparison,rappel2017mechanisms}, cytoskeletal dynamics \cite{huang2013excitable}, cell deformation \cite{shi2013interaction} and wave-like propagation of sub cellular level components \cite{beta2017intracellular}.

%Independently of the presence of an external signal, before moving, cells need to define which direction to follow. To do so, they first need to define one front and one back. This process is commonly known as cell polarization \cite{mogilner2012cell}. 
Once the axes and direction of movement are defined, in amoeboid cells small projections (defined as protrusions) are formed in the membrane of the cells \cite{bosgraaf2009ordered}. 
These projections, which extend and retract periodically, are responsible for pushing the cell to move in the typical amoeboid motion. \textit{Dictyostelium discoideum} is a species of soil-dwelling amoeba, frequently employed to characterize amoeboid locomotion. Inside the projections of this amoeba, several signaling events are triggered, for example the activation of the Ras proteins and PI3K enzymes and accumulation of $PIP_3$ at the front of the cell, while activation of PTEN and myosin occur at the rear of the cell \cite{kae2004chemoattractant, etzrodt2006time}. Membrane areas where the protrusion activity is greater are typically characterized by the presence of Ras-GTP protein zones, denoted as patches \cite{sasaki2007g}. The appearance of Ras regions around the membrane with high protrusion activity \cite{xiong2010cells,van2017coupled,shi2013interaction} has been observed in the slime mold organism \textit{Dictyostelium discoideum} where it is related to cytoskeletal dynamics \cite{huang2013excitable}. % and is responsible for cell deformations \cite{shi2013interaction}.

%In the well-studied slime mold organism \textit{Dictyostelium discoideum} it has been shown the appearance of Ras regions around the membrane with high protrusion activity \cite{xiong2010cells,van2017coupled}.

 A reaction-diffusion model with bistable dynamics is one of the common models of cell motility and particularly employed for \textit{Dictyostelium discoideum}. Some dimensional bistable models are based on the formation of a finite lifetime and localized patches of high protein concentration \cite{hecht2010transient, hecht2011activated} while others are based on membrane dynamics of moving connected points \cite{neilson2011chemotaxis}. Bistable conditions for cellular processes can be obtained by the coupling of a mass control regulating condition in the biochemical components present around the cytosol and the membrane of the cell, for example proteins, phospholipids and enzymes present or other substances in the cytoskeleton \cite{halatek2018rethinking,mori2008wave,otsuji2007mass}.

A common technique to model pattern formation inside a cell and the shape evolution of the membrane is the addition of a phase field with a sharp interface to distinguish the interior and exterior of the cell. This field maintains the correct boundary conditions while the borders are moving \cite{camley2013periodic}. Some studies have applied phase field modeling to study keratocyte motility \cite{shao2012coupling,camley2017crawling,nickaeen2017free} and amoeboid motility \cite{alonso2018modeling} which can be divided into diffuse and persistent migration depending on the starvation level \cite{stankevicins2020deterministic,ecker2021excitable}. There are also some intermediate cases observed in  \textit{Dictyostelium discoideum} cells for certain types of genetic variants \cite{miao2017altering} which have also been modelled with a phase field and variation of some parameters \cite{moreno2020modeling,cao2019minimal,cao2019plasticity}. Other properties of \textit{Dictyostelium discoideum} cells
modeled employing a phase field are the viscoelasticity of the cells \cite{moure2017phase} and cell division \cite{flemming2020cortical}.
Finally, we would like to mention that  interactions among cells can also be considered for keratocyte \cite{lober2015collisions,kulawiak2016modeling}  and amoeboid cells \cite{moreno2021}.

%Here, using a reaction diffusion model 
Here, first, we transform a one dimensional model of the polarization at the membrane of a single \textit{Dictyostelium discoideum} cell \cite{van2017coupled} into a two-dimensional domain for the waves in the basal membrane, in contact with the surface, using an additional phase field for the shape of the cell. 
We observe a strong dependence of the numerical dynamics on the explicit parameter values of the original model and on the intensity of the stochastic fluctuations. Such dependence indicates that the model is not robust to small variations of the parameters.
Therefore, we propose a constraint on the conservation of a component of the signal pathway, controlling the autocatalytic mechanism, in order to increase the robustness of the model in representing changes in parameter values \cite{kitano2007towards}. 
Similar types of conservation have previously been employed in other models of motion of  \textit{Dictyostelium discoideum} cells \cite{alonso2018modeling,moreno2020modeling,imoto2020comparative,taniguchi2013phase}, and we show here that a constraint of this kind is an useful and reasonable condition to systematically increase the robustness of the crawling mechanism, increasing the window of parameter values allowing cell migration.

%Patches and wave-like propagation of sub cellular level components have been modeled using reaction-diffusion systems

%This direction can or can not be related with the presence of an external signal. 

\section{Model and Methods}

\subsection{Biochemical model for Ras activation and pseudopod extension} %extension

Biochemical components inside the \textit{Dictyostelium discoideum} amoeba self-organized to follow chemical signals in the exterior by the accumulation of Ras-GTP in the front of the cell. Next, F-actin molecules accumulates also in the front triggering push of the cytoskeleton and the motion of the amoeba.   

\begin{figure}[h!]
\begin{center}
\includegraphics[width=\linewidth]{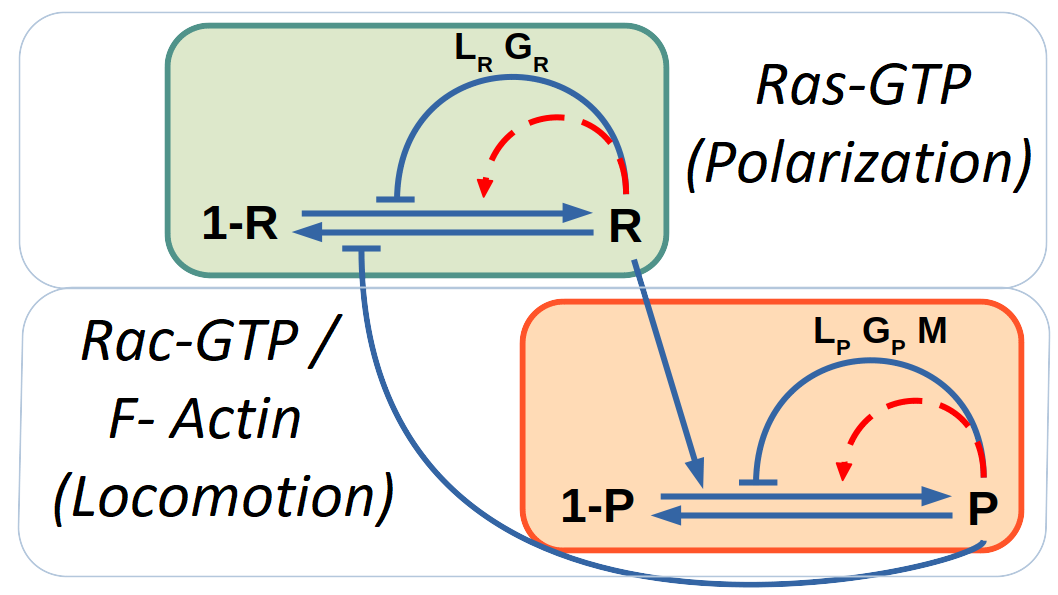}% This is a *.eps file
\end{center}
\caption{Sketch of the biochemical model of polarization and locomotion modules. Blue solid lines corresponds to the interactions and red dashed lines to auto-catalytic processes. The mass-conservation constrains control the auto-catalytic interactions.}
  \label{fig:fig0}
\end{figure}

We investigate a simple reaction-diffusion model \cite{van2017coupled} for Ras-GTP (R) patches that consists of a local activator and a global inhibitor enhanced by the activation of the Ras-GTP; see diagram in Fig.~\ref{fig:fig0}. 
The partial differential equations employed are:

%\cite{van2017coupled}

\begin{eqnarray}
\label{eqn:equno}
\frac{\partial R}{\partial t}&=(1-R)(k_1+k_2\frac{R^{n1}}{R^{n1}+K_R^{n1}}-k_3G_R-k_4L_R)
\nonumber\\ 
%-k_5R\frac{1}{1+k_6P}+D_R\nabla^2R+N_R(0,\sigma_R)  
&-k_5R\frac{1}{1+k_6P}+D_R\nabla^2R+ \xi_R(\vec{x},t),
\end{eqnarray}

\begin{equation}
\label{eqn:eqdos}
\frac{\partial G_R}{\partial t}=k_7\langle R \rangle - k_9G_R,
\end{equation}

\begin{equation}
\label{eqn:eqtres}
\frac{\partial L_R}{\partial t}=(1-L_R)k_{10}R-k_{11}L_R+D_{LR}\nabla^2 L_R;
\end{equation}
where $\xi_R(\vec{x},t)$ is a Gaussian spatio-temporal distributed white noise with zero mean $ \langle \xi_R(\vec{x},t) \rangle = 0 $ and correlation  $ \langle \xi_R(\vec{x},t) \xi_R(\vec{x'},t') \rangle = 2 \sigma_R \delta(\vec{x}- \vec{x'}) \delta(t-t') $. The variables $L_R$ and $G_R$ correspond, respectively, to local and global inhibitors of R. Finally, note that the quantity $\langle R \rangle$ corresponds to the spatial integration in the whole system of the field R.

At the same time we included a quantity related to the formation of protrusions (P) such as F-actin and Rac-GTP molecules; see diagram in Fig.\ref{fig:fig0}. 
These variables were coupled with their respective inhibitors:

\begin{eqnarray}
\label{eqn:eqcuatro}
\frac{\partial P}{\partial t}&=(1-P)(k_{12}+k_{13}R+k_{14}\frac{P^{n2}}{P^{n2}+K_P^{n2}}
\\ \nonumber
&+k_{15}\frac{M^{n3}}{M^{n3}+K_M^{n3}}-k_{16}G_P-k_{17}L_P)-k_{18}P
\\ \nonumber
%&+D_P\nabla^2P+N_P(0,\sigma_P)
&+D_P\nabla^2P+\xi_P(\vec{x},t),
\end{eqnarray}

\begin{equation}
\label{eqn:eqcinco}
\frac{\partial G_P}{\partial t}=k_{19}\langle P \rangle - k_{20}G_P,
\end{equation}

\begin{equation}
\label{eqn:eqseis}
\frac{\partial L_P}{\partial t}=(1-L_P)k_{21}P-k_{22}L_P+D_{LP}\nabla^2 L_P;
\end{equation}
where $\xi_P(\vec{x},t)$ is a Gaussian spatio-temporal distributed white noise with zero mean $ \langle \xi_P(\vec{x},t) \rangle = 0 $ and correlation  $ \langle \xi_P(\vec{x},t) \xi_P(\vec{x'},t') \rangle = 2 \sigma_P \delta(\vec{x}- \vec{x'}) \delta(t-t') $. As in the previous set of equations, the variables $L_P$ and $G_P$ correspond, respectively, to local and global inhibitors of P; while the quantity $\langle P \rangle$ corresponds to the spatial integration in the whole system of the field P.

Finally, the model takes into account the inclusion of a variable of memory (M) which is coupled with P; this variable stimulates the formation of new protrusion zones and represents the results observed in some experiments where a certain relation between the formation of new pseudopods in previous pseudopod locations is present \cite{cooper2012excitable,van2021short}:
\begin{equation}
\label{eqn:eqsiete}
\frac{\partial M}{\partial t}=k_{23}P-k_{24}M+D_{M}\nabla^2 M.
\end{equation}
The memory term is related to the probability of identifying when and where the signaling cascade is activated to generate new pseudopods. At a molecular level, such term is identified with the mechanisms of how information is collected, stored and used to bias future pseudopods \cite{van2021short}.

For a more exhaustive description of the model, check the original study \cite{van2017coupled}. Note that the noise description in the original study was different and we have adapted to an equivalent description based on physical derivations of stochastic fluctuations \cite{garcia2012noise}; for more details see appendix A.
%sigma^2 = variance = .04
%sigma = standard dev = .04

%From equations 1 and 4 which correspond to the Ras activation and pseudopod formation, respectively we can notice that both contain a noise term $N(0,\sigma)$, where that noise is a Gaussian white noise with mean $0$ and variance $\sigma^2$. The inclusion of noise for both R and P generates a different outcome of the model dyanamics for every simulation.

%Desviacion estandar es $\sigma=0.04$

One dimensional simulations were made using periodic boundary conditions and using 120 points for the grid, see an example in Fig.(\ref{fig:fig1p5},A) where R and P are plotted as function of time. The cell was considered as circular and having a radius of $6.25$ $\mu m$. The pixel size for this case was set at $\Delta x=0.32$ $\mu m $ and the time step $\Delta t=0.03s$
The definition and the value of the parameters of the model can be found in Table 1 in appendix B.
%\begin{equation}
%\resizebox{0.75\textwidth}{!}{$\frac{\partial R}{\partial t}=(1-R)(k_0+k_1S+k_2\frac{R^{n1}}{R^{n1}+K_R^{n1}}-k_3G_R-k_4L_R)-k_5R\frac{1}{1+k_6P}+D_R\nabla^2R+N_R(0,\sigma_R)$}
%\end{equation}

\subsection{Physical phase field model for cell shape deformations}

\begin{figure}[ht!]
\begin{center}
\includegraphics[width=\linewidth]{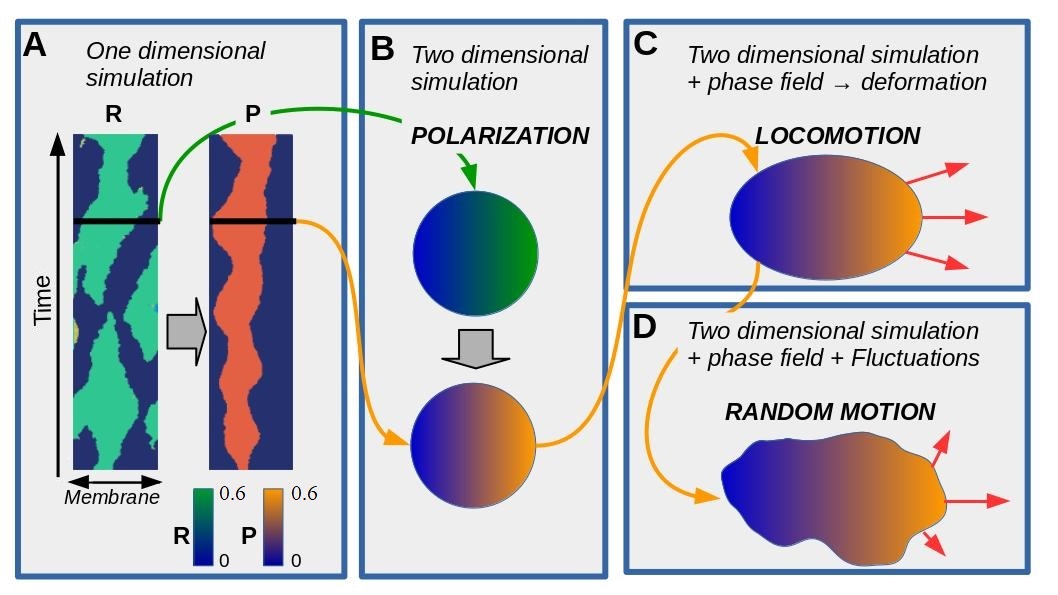}% This is a *.eps file
\end{center}
\caption{Explanatory box of the modelling techniques. A) Kymograph of a typical one dimensional simulation showing the formation of patches of R and P (horizontal direction corresponds to position in the membrane, and vertical direction to the time evolution ). B) Extension to two dimensions with the addition of a circular phase field (Value 1 inside and 0 outside of the circular border). Patches of R and P form in the basal surface of the cell. C) Addition of a dynamic phase field producing the motion and the deformation of the cell. D) Addition of the fluctuations, random two dimensional patches are distributed inside the phase field, driving the motion of the cell.}
  \label{fig:fig1p5}
\end{figure}

The original model did not include deformable cells. Therefore, we expanded the model to 2D geometry, see  Fig.(\ref{fig:fig1p5},B) for a simple extension, and coupled it to an auxiliary phase field $\phi$ with the purpose of describing the evolution of the cell shape. The use of a phase field permits a smooth variation between the values of $\phi=1$ inside and $\phi=0$ outside of the cell, respectively. See an example in Fig.(\ref{fig:fig1p5},C) where the field P deforms the shape. Note that the fluctuations in P will also affect the shape, see Fig.(\ref{fig:fig1p5},D).

The phase field equation is the result of a force balance involving several types of forces of different natures acting in the cell body. The equation for the phase field is as follows

\begin{eqnarray} \label{eqn:pf}
\tau \frac{\partial \phi}{\partial t} &=& \gamma \left(\nabla^2 \phi -\frac{G'(\phi)}{\epsilon^2}\right)
\\  \nonumber
& & - \beta \left(\int \phi\, dA - A_0 \right)\left| \nabla \phi \right| 
+\alpha\, \phi\, P \left| \nabla \phi \right|  \,, 
\end{eqnarray}
The first term on the right side in eq. (\ref{eqn:pf}) is related to the surface energy of the cell membrane, where $\gamma$ is the surface tension and $G(\phi)=18\phi^2(1-\phi)^2$ is a double well potential.The second term helps keep the area close to the value of $A_0$. And the third term represents the active force generated by the Rac-GTP (P) molecules when pushing on the cell membrane \cite{alonso2018modeling}. 

The inclusion of the phase field as previously mentioned implies a change of geometry from 1D to 2D. Here we consider a deformable cell of $122 \mu m^2$ in area corresponding to a circle with radius equal to $6.25\mu m$. The pixel size used was the half as in the 1D case $\Delta x=\Delta y=0.16 \mu m$. Also, to increase accuracy, the time discretization was reduced to $\Delta t=0.003s$. 
%Different sizes for the square grid were used, being $300\times300$, $400\times400$ and $500\times500$. Finally 
We integrated Eqs. (\ref{eqn:equno}-\ref{eqn:pf}) using periodic boundary conditions and standard finite differences. The rest of parameters are as displayed in Table 1 in appendix B.

%From the simulations different cell phenotypes, trajectories and velocities were obtained 
% YA he referenciado la figura arriba Eduardo 
%In Figure \ref{fig:fig1p5} is shown a more detailed explanation of the modeling techniques used in the development of this work.

\subsection{Mathematical description of the  mass-conservation constraint}

%The mathematical model described in Eqs(1-7) has several conservation terms which affect the dynamics of the system, see for example Eq.(5) and Eq.(7). However, we include a feedback control through the parameter k$_2$ to control the bistability shown by the model, see next sections. Therefore, the parameter k$_2$ is dynamically controlled depending on the total amount of the protein R at the membrane (O ALGO ASI¿?):

%\begin{eqnarray} \label{eqn:k2}
%k_2 &=&  k_{2_0} + \left(  \langle R \rangle -C_0 \right)  \,, 
%\end{eqnarray}

The mathematical model represented in Eqs. (\ref{eqn:equno}-\ref{eqn:eqsiete}) has several conservation terms which affect the dynamics of the system; see for example Eq.(\ref{eqn:eqdos}) and Eq.(\ref{eqn:eqcinco}). 

However, we modify the original model to include two new feedback through parameters $k_2$ and $k_{14}$ to create a more robust model by the control of the bistability properties of the model, which are discussed in following sections. Therefore, the parameters $k_2$ and $k_{14}$ are dynamically controlled depending on the total amount of the protein R on inducer P at the cell, respectively.

The new control term related to $k_2$ is  
\begin{eqnarray} \label{eqn:k2}
k_2 &=&  k_{2}^* + \eta \left(  \langle R \rangle -C_R \right)  \,, 
\label{eqk2}
\end{eqnarray}
Also, in a similar way, the control term for $k_{14}$ is as follows:
\begin{eqnarray} \label{eqn:k14}
k_{14} &=&  k_{14}^* + \eta \left(  \langle P \rangle -C_P \right)  \,, \label{eqk14}
\end{eqnarray}
where $k_2^*$ and $k_{14}^*$ are new constants that take the values of the original value of $k_2$ and $k_{14}$, respectively. Note that for $\eta$=0 we recover the original model. 

Parameter $C_R$ is the target value of the fraction of the cell area occupied by patches of Ras $R$. When, the total concentration $ \langle R \rangle$ is larger (lower) than $C_R$ the parameter value of $k_2$is larger (lower) than $k*_2$ and the bistable conditions implies the reduction (increase) of the R and, therefore, of $ \langle R \rangle$. It corresponds to a feedback control of the parameter to keep a particular stable solution in Eq.(\ref{eqk2}). The parameter $C_P$ is the target fraction occupied by the pseudopod inducer $P$ and produces an equivalent dynamics on Eq.(\ref{eqk14}).

\section{Results}

\subsection{Mechanism of cell motion in the mathematical model is based in stochastic generation of patches}

In a one-dimensional system, mimicking the membrane of a crawling cell, the generation of a local patch of high biochemical concentration is equivalent to cell polarization. 
A stable domain in a specific location of the membrane is related with actin accumulation and produces persistent motion of the cell in that direction. 
The random appearance and disappearance of small domains is related with amoeboid motion, where two or three pseudopods compete for a certain time. The alternation of direction gives rise to random motion of the virtual cell.

\begin{figure}[h!]
\begin{center}
\includegraphics[width=\linewidth]{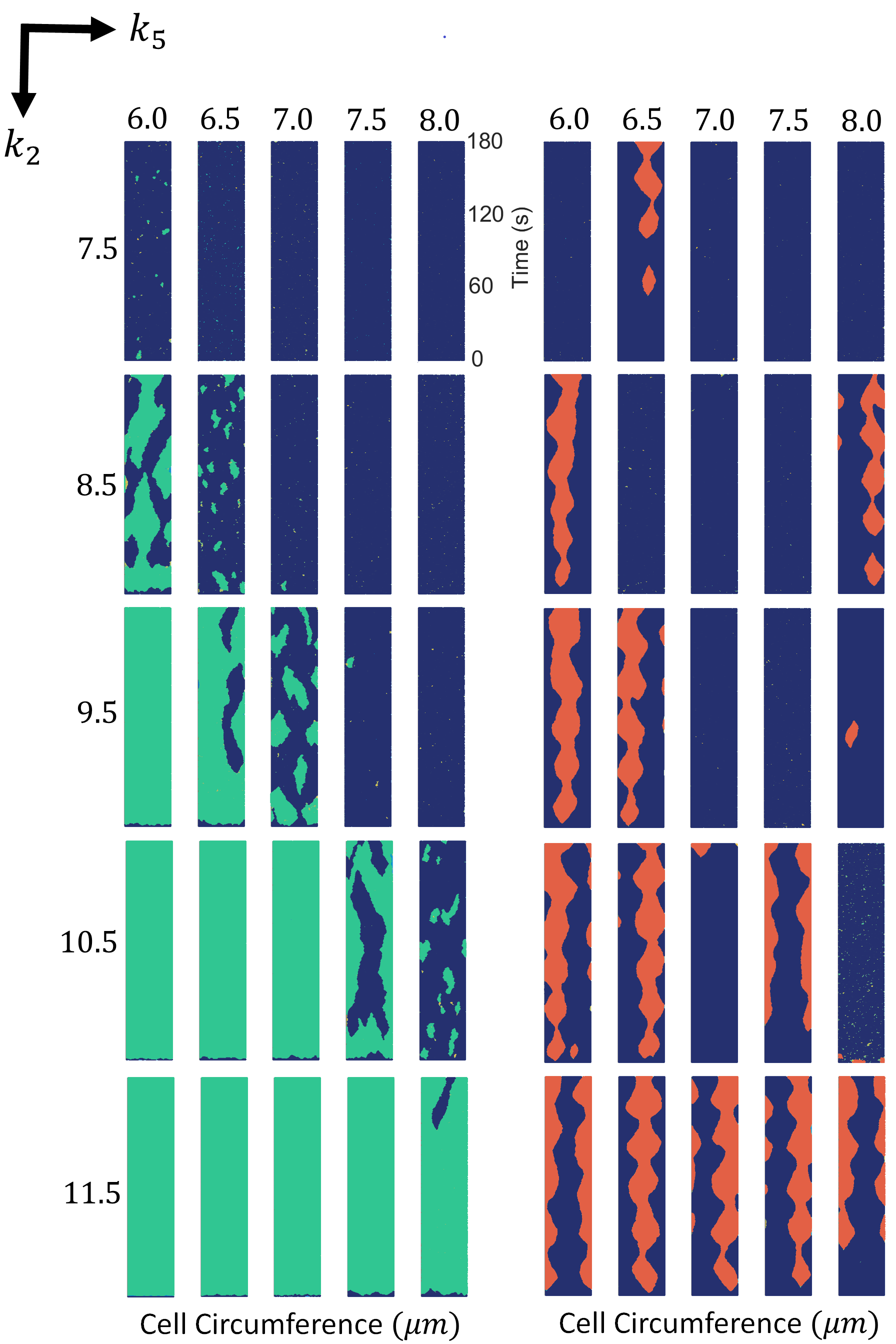}% This is a *.eps file
\end{center}
\caption{Spatio-temporal plots of R (in green) and P (in red) for different values of $k_5$ and $k_2$. For each panel times goes from bottom to top and horizontal direction corresponds to membrane position in the cell circumference. Variances of noise intensity were fixed at $\sigma_R=0.04$ and $\sigma_P=0.025$.}
  \label{fig:fig1}
\end{figure}

Stochastic reaction-diffusion equations, as in Eqs.( \ref{eqn:equno}-\ref{eqn:eqsiete}), and previously developed \cite{van2017coupled}, can be numerically integrated into a one-dimensional domain as in the original study. Kymographs are shown next for a certain window of parameter values, the spatio-temporal dynamics of concentration R displays relatively stable patches which appear at different locations; see spatio-temporal plots in Fig.~\ref{fig:fig1}.

In the simulations shown in Fig.~\ref{fig:fig1}, the parameters $k_{14}$ and $k_{18}$ are kept constant and the response of P along the membrane is quite similar for the parameters considered. There are some persistent patches during a long period of the simulations. Under such conditions the cell may develop its own locomotion. 
Note that the activation of P by R is weak; a strong coupling, controlled by parameter $k_{13}$, would correlate the patches of both fields. We keep the original values of this parameter \cite{van2017coupled}.
For small values of the parameter $k_5$ and large values of $k_2$, the concentration of R is high and homogeneously distributed along the membrane; see corresponding spatio-temporal plots in Fig.\ref{fig:fig1}. The concentration of R increases the probability of the generation of patches of P, and therefore, depending on the particular realization, some direct motion is expected under such conditions. 
%Although  whole membrane is covered with the protein related with R the coverage of P may give rise to direct motion. 
On the other hand, in the opposite limit, large values for parameter $k_5$ and small values for $k_2$, the concentration of R is low, and the probability of directed motion is smaller. 
%In between, for intermediate values of k$_2$ and k$_5$, see values in Fig.\ref{fig:fig1}, patches are formed and are relatively stable, and persistent motion is expected. 

%However, the parameters $k_{14}$ and $k_{18}$ are kept constant during the previous simulations and the response of P along the membrane is quite similar for the parameters considered. There are some persistent patches during a large time of the simulations. Under such conditions actually independently on the dynamics of R the cell may develop its own locomotion. Note that if the activation of P by R, controlled by parameter k$_{13}$, is strong, it will strongly correlate the patches of both fields. We keep the original values of this parameter \cite{van2017coupled}.

\begin{figure}[h!]
\begin{center}
\includegraphics[width=\linewidth]{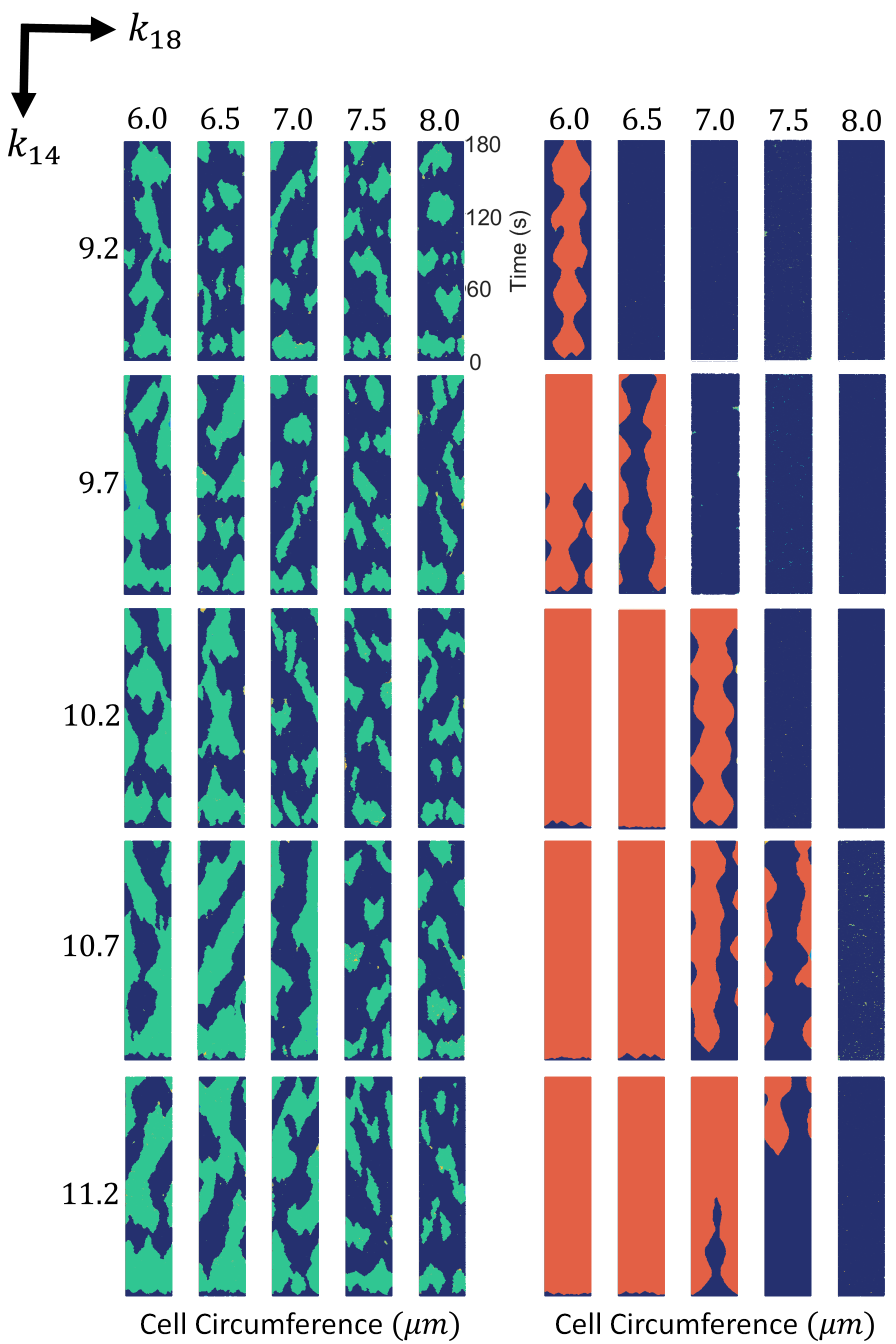}% This is a *.eps file
\end{center}
\caption{Spatio-temporal plots of R (in green) and P (in red) for different values of $k_{14}$ and $k_{18}$. For each panel times goes from bottom to top and horizontal direction corresponds to membrane position in the cell circumference. Variances of noise intensity were fixed at $\sigma_R=0.04$ and $\sigma_P=0.025$.}
  \label{fig:fig2}
\end{figure}

If we fix $k_{2}$ and $k_{5}$ values and vary the parameters controlling for the evolution of P, we obtain the inverse behaviour; see Fig.\ref{fig:fig2}. For large values of $k_{14}$ and small values of  $k_{18}$ the membrane is completely covered by P, while for small values of $k_{14}$ and large values of $k_{18}$ the concentration of P strongly decreases. Locomotion is, therefore, expected for intermediate values of these two parameters as shown in   Fig.~\ref{fig:fig2}. However, the effect of changing such parameters on the dynamics of R is small.

\subsection{Mechanism of amoeboid motion is based in bistability}

\begin{figure}[h!]
\begin{center}
\includegraphics[width=\linewidth]{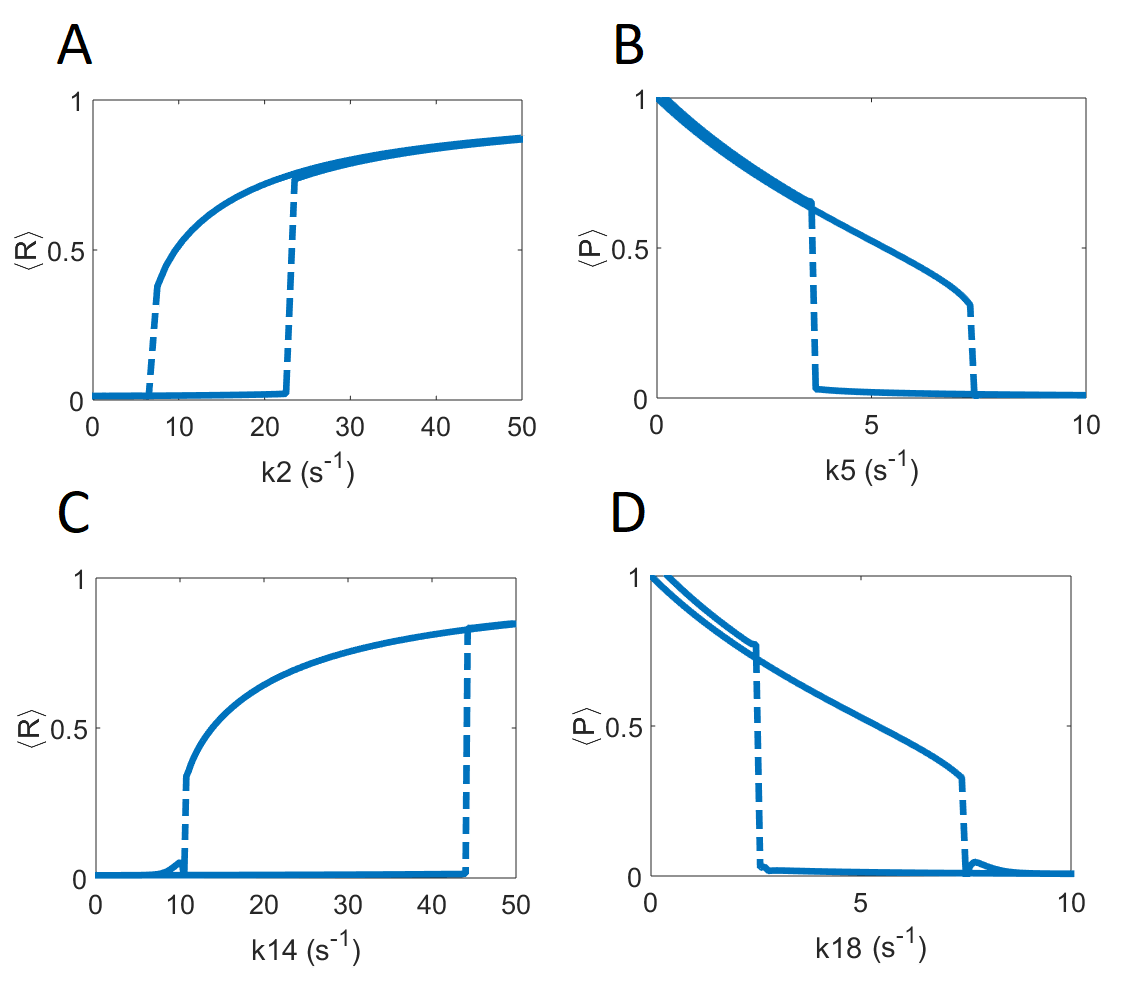}% This is a *.eps file
\end{center}
\caption{Hysteresis curves of the deterministic model for R dynamics varying (A) $k_2$ and (B) $k_5$. Similar hysteresis curves of the deterministic model for P dynamics were obtained when varying (C) $k_{14}$ and (D) $k_{18}$. }
  \label{fig:fig3}
\end{figure}

For the evaluation of the mechanism of the spatio-temporal plots in Figs.~\ref{fig:fig1} and \ref{fig:fig2}, we fixed the stochastic source of perturbations of the additive noises in eqs (\ref{eqn:equno}-\ref{eqn:eqsiete}) to $\sigma_R=0.04$ and $\sigma_P=0.025$. No changes were expected in the average behaviour or in the quantification the value of the total concentration R in the face of the parameters $k_2$ and $k_5$ because the fluctuations are additive \cite{garcia2012noise}. 
Stationary solutions were found for the deterministic version of Eqs. (\ref{eqn:equno}-\ref{eqn:eqsiete}). Slowly increasing the parameter $k_2$ we observed the evolution of the stationary value of R; see Fig.~\ref{fig:fig3}. Once the system of equations saturated to a certain value of R$ \sim 0.85-0.90$ we reduced the value of $k_2$, giving rise to a hysteresis cycle; see Fig.~\ref{fig:fig3}A. The hysteresis cycle shows that the system actually follows bistable dynamics. 
Equivalent dynamics of R can be observed under a similar change in $k_5$; see Fig.~\ref{fig:fig3}B. 

In contrast, the same analysis was made for the constants $k_{14}$ and $k_{18}$, which are related to the variable of the pseudopod inducer P. The resulting dynamics revealed bistable behavior for two stable states as shown in the hysteresis curve for parameters $k_{14}$ (Fig.~\ref{fig:fig3}C) and $k_{18}$ (Fig.~\ref{fig:fig3}D).

It is known that the combination of a bistable dynamics with an appropriate noise intensity, produces the formation of localized patches in reaction-diffusion equations \cite{hanggi1985bistability,kramers1940brownian}. This formation is the mechanism responsible for the formation of localized domains of R and P. The localized pattern observed in Fig.~\ref{fig:fig1} is not due to excitable dynamics but rather to bistable dynamics with noise.

\subsection{Stochastic bistability provides a mechanism of cell crawling motion in two dimensional systems}

For the coupling of the polarization mechanism to the proper cell motion shown in the previous section we used a phase field. This additional field is employed for the definition of the interior of the cell. In this case, the two dimensional phase field permits the use of different biochemical concentrations in the ventral part of the cell in a surface. 

\begin{figure*}[t!]
\begin{center}
\includegraphics[width=\linewidth]{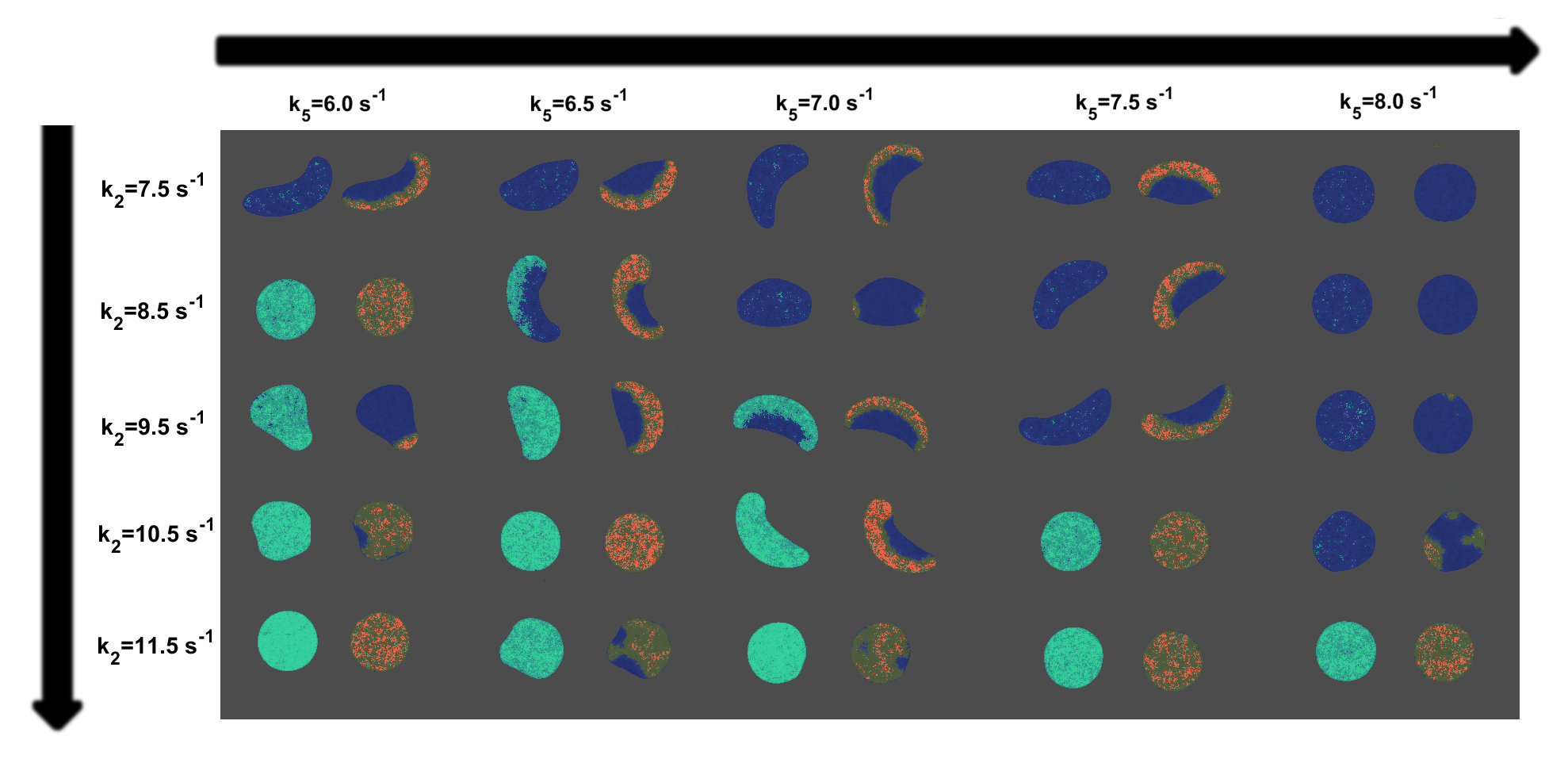}% This is a *.eps file
\end{center}
\caption{Snapshots of the virtual cells showing R (in green) and P (in red) obtained from computer simulations and applying the phase field technique to the model for different values of $k_2$ and $k_5$. Variances of noise intensity were fixed at $\sigma_R=0.04$ and $\sigma_P=0.025$. }
  \label{fig:fig4}
\end{figure*}

The extension of the stochastic reaction-diffusion described in the previous section to two dimensions permits numerical simulations of the shape of the crawling cell and the corresponding motion responding to the dynamics of the patches. In Fig.~\ref{fig:fig4} we see snapshots of the \textit{in silico} cells with different parameter values corresponding to equivalent parameter values shown in Fig.~\ref{fig:fig1}. 
We reproduced the dynamics expected from the one dimensional simulations and the crawling dynamics can be clearly studied. Again, for small values of parameter $k_5$ and large values of $k_2$ the concentration of R is mostly high and homogeneously distributed along the membrane; see Fig.~\ref{fig:fig4}. On the other hand, in the opposite limit, for large values of $k_5$ and small values of $k_2$ the concentration of R disappears from the ventral membrane. However, for a small window of values of the parameters the cell moves and inspects the surrounding region. The shape of the resulting cell depends on the particular realization and the parameter values; in the snapshots shown in Fig.~\ref{fig:fig4} we obtain fan-shaped cells, a typical mode of \textit{Dictyostelium discoideum} cell crawling \cite{miao2017altering}.  

%The velocity of the cellular motion XXXX. (????)

%%%%PARRAFO ANTES
%From the snapshots results from Fig.~\ref{fig:fig4} only three scenarios show persistent movement together with correlation in the concentrations R and P. 
%They correspond to the couple of values: $(k_2,k_5)={(8.5,6.0),(9.5,7.0),(10.5,8.0)}$ (units of $s^{-1}$). 
%The model exhibits sensitivity to variations of the parameters k$_2$ and k$_5$. 
%Changes on the values brings the cell from a quite persistent movement into an alternation between almost static state for low concentrations of <R>, and transitory motions in random directions, for large values of <R>. As in Fig.\ref{fig:fig1} the influence of R in P is evident in these two-dimensional simulations. Note that the patches of P appearing for low values of k$_2$ are not stable, and not strong persistent motions are observed. 
%%HASTA AQUI

From the snapshots resulting from Fig.~\ref{fig:fig4} different scenarios show different types of movement together with the concentrations of R and P. 
%ESTE ANTES NO ESTABA COMENTADO They correspond to the couple of values: $(k_2,k_5)={(8.5,6.0),(9.5,7.0),(10.5,8.0)}$ (units of $s^{-1}$). 
%Such results implies that t
The model exhibits sensitivity to variations in parameters $k_2$ and $k_5$. 
Different realizations of simulations with the same parameters could bring the cell from quite persistent movement into an alternation between an almost static state for low concentrations of $\langle R \rangle$, and transitory motion in random directions for large values of $\langle R \rangle$. As in Fig.\ref{fig:fig1} the small influence of R in P is evident in these two-dimensional simulations. Note that the presence of patches of P appearing for high values of $k_2$ and low values of  $k_5$ is more common because $\langle R \rangle$  slightly increases the probability of forming new $\langle P \rangle$ patches. In contrast, for low values of $k_2$ and high values of  $k_5$ the presence of $\langle P \rangle$ patches is less frequently observed. %and not strong persistent motions are observed. 

%NEW FIGURE LIKE 5

\begin{figure*}[t!]
\begin{center}
\includegraphics[width=\linewidth]{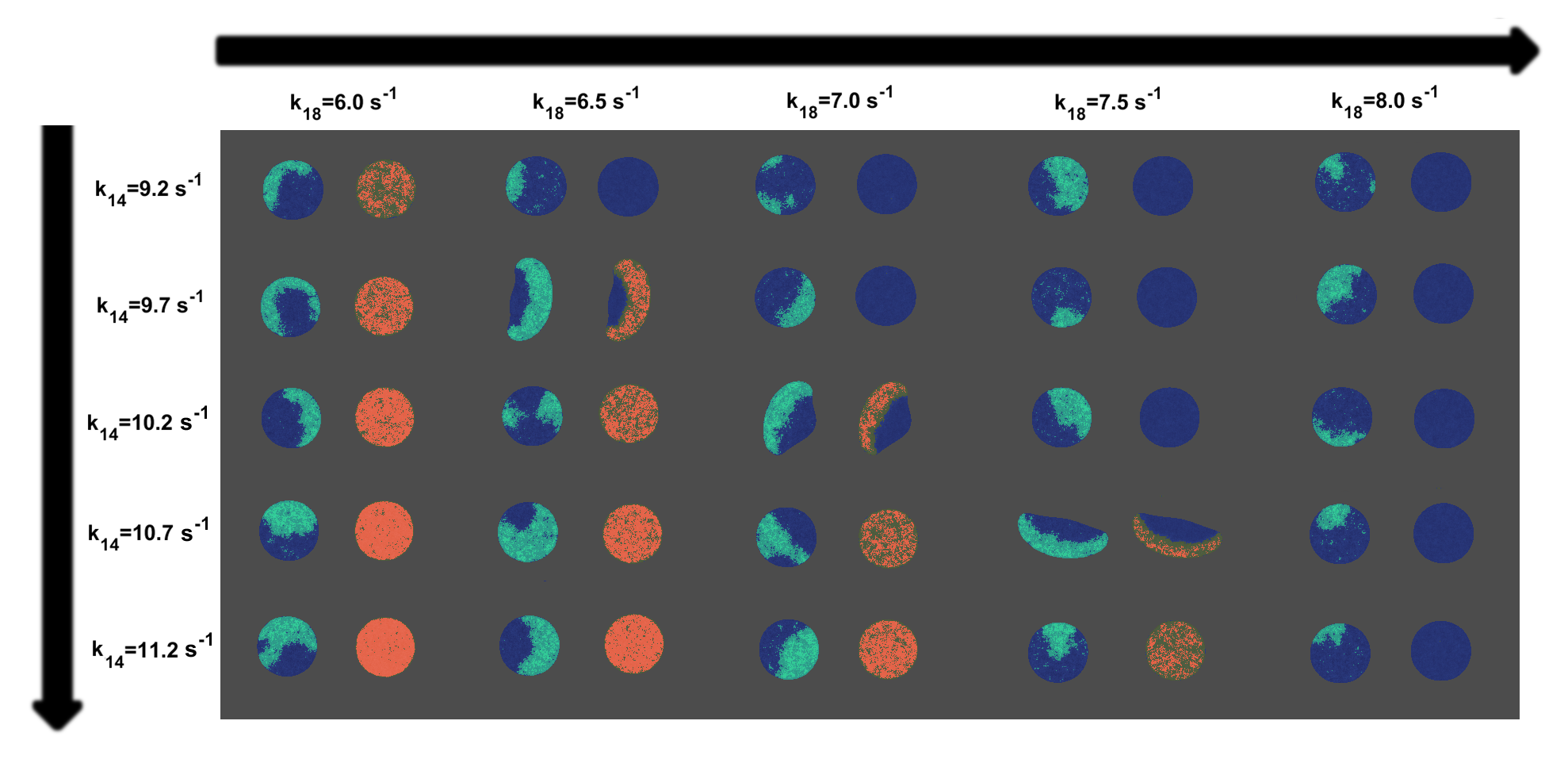}% This is a *.eps file
\end{center}
\caption{Snapshots of the virtual cells showing R (in green) and P (in red) obtained from computer simulations and applying the phase field technique to the model for different values of $k_{14}$ and $k_{18}$. Variances of noise intensity were fixed at $\sigma_R=0.04$ and $\sigma_P=0.025$. }
  \label{fig:fig4.5}
\end{figure*}

In contrast to this, the dependence of locomotion on the parameters $k_{14}$ and $k_{18}$ is much stronger as shown in Fig.\ref{fig:fig4.5}, where persistent motion is observed only for intermediate values of the two parameters. For very small or very large amounts of $\langle P \rangle$ cells do not move, independently of the behaviour of R.    

\subsection{Noise intensities determine the type of motion}

\begin{figure*}[t!]
\begin{center}
\includegraphics[width=\linewidth]{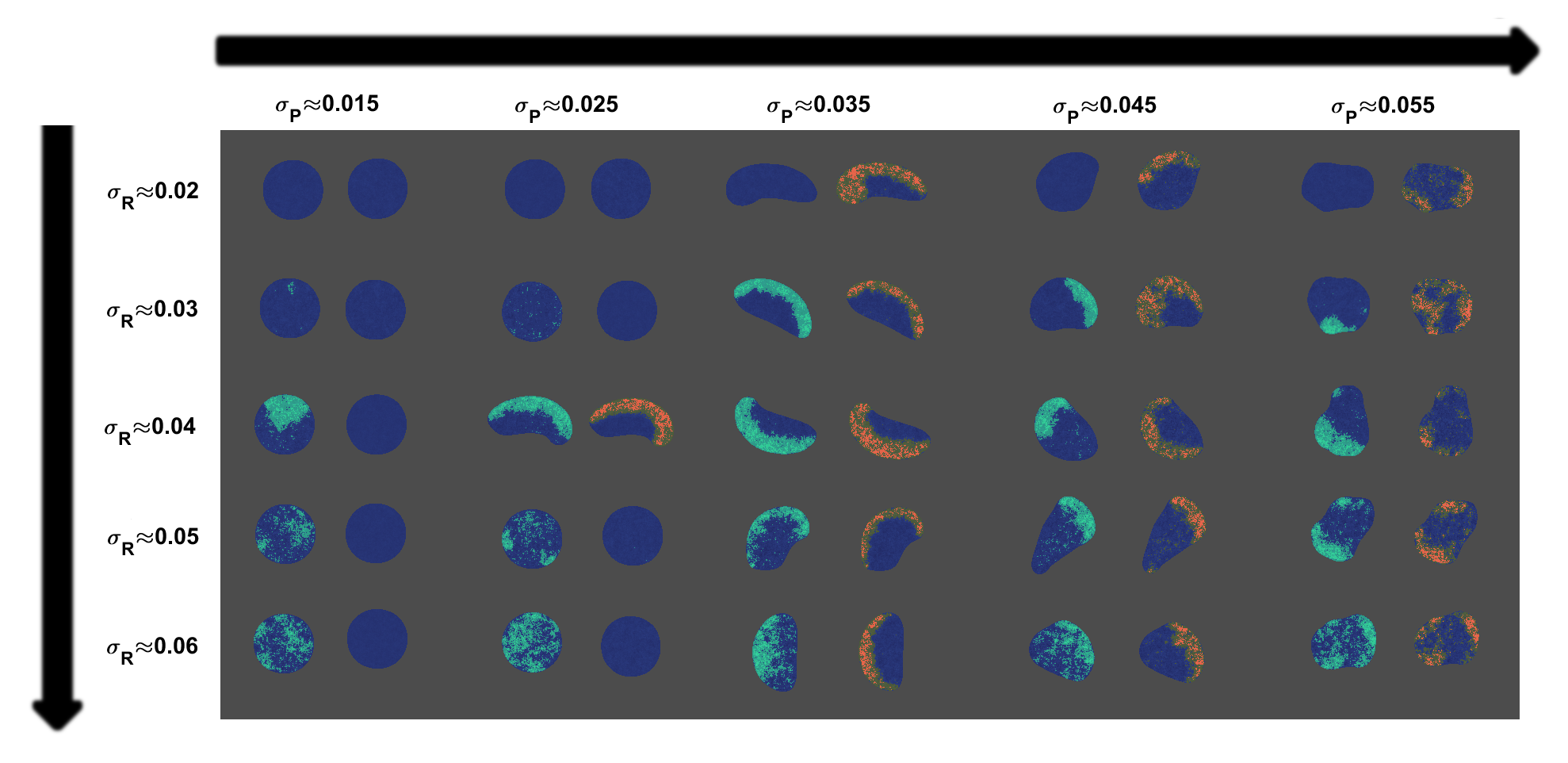}% This is a *.eps file
\end{center}
\caption{Map of snapshots of the virtual cell taken from different variance values of noise intensity for the original model \cite{van2017coupled} . %Results obtained correspond to
%0.025 y 0.04 son los valores estandar del modelo de VanHaastert, solo fui variando el ruido de P y R, es decir cuando se quedo fijo P cambie R y cuando se quedo fijo R se cambio P. 
For every column, left/green snapshots refer to R distribution while right/red snapshots indicate P distribution. The value of the parameter of the simulations are $k_{2}=9.50$, $k_{5}=7.0$, $k_{14}=10.20$ and $k_{18}=7.0$. The rest of them are consistent with Table 1. }
  \label{fig:fig5}
\end{figure*}

The variation of the type of cell motion is determined by the intensity of the noise. Low levels of both noises do not permit the formation of the domains of R and P needed to give rise to cell movement; see Fig.\ref{fig:fig5}. 
As we increase the amplitude of noise for P, we reach a single domain that fills the front part of the cell and persistent motion is shown; see middle column and right hand snapshots in Fig.\ref{fig:fig5}. 
This type of motion is reminiscent of the fan-shaped amoeboid cells previously reported \cite{miao2017altering} and also reproduced with simpler models \cite{moreno2020modeling}. 
An increase in the noise intensity produces an increase in the appearance and disappearance of pseudopods and therefore a transition to amoeboid movement; see fifth column in Fig.\ref{fig:fig5} for higher noise intensities. 
Another evident change the effects of which will be discussed in the following sections is the variance associated with the formation of Ras patches ($\sigma_R$) becoming varied. %Although we will discuss this effect in the next sections. 
There are some noise intensities which produce patterns and dynamics similar to the characteristic patterns observed in the experiments, however, as previously mentioned, and shown in Fig.\ref{fig:fig7}, small variations in the parameter values may completely change the movement dynamics.

%UN RUIDO de P en 0.06 generara pseudopods mas aleatorios y si junto con esta condicion ponemos ruidos menores de R, por ejemplo 0.04 se generaran patches de R mejor formados pero que iran apareciendo y desapaerciendo en diferentes zonas dependiendo de P

\subsection{Inclusion of mass-conservation stabilizes cell motion}

In order to reduce the high sensitivity of the movement on the parameter values, we included a conservation constraint on the total quantities of the protein Ras R and the inducer P in the biochemical rates responsible for the bistable transition. Following similar previous approaches \cite{alonso2018modeling,imoto2020comparative}, we included this conservation as a global feedback condition in the reaction-diffusion equations. 

\begin{figure*}[t!]
\begin{center}
\includegraphics[width=\linewidth]{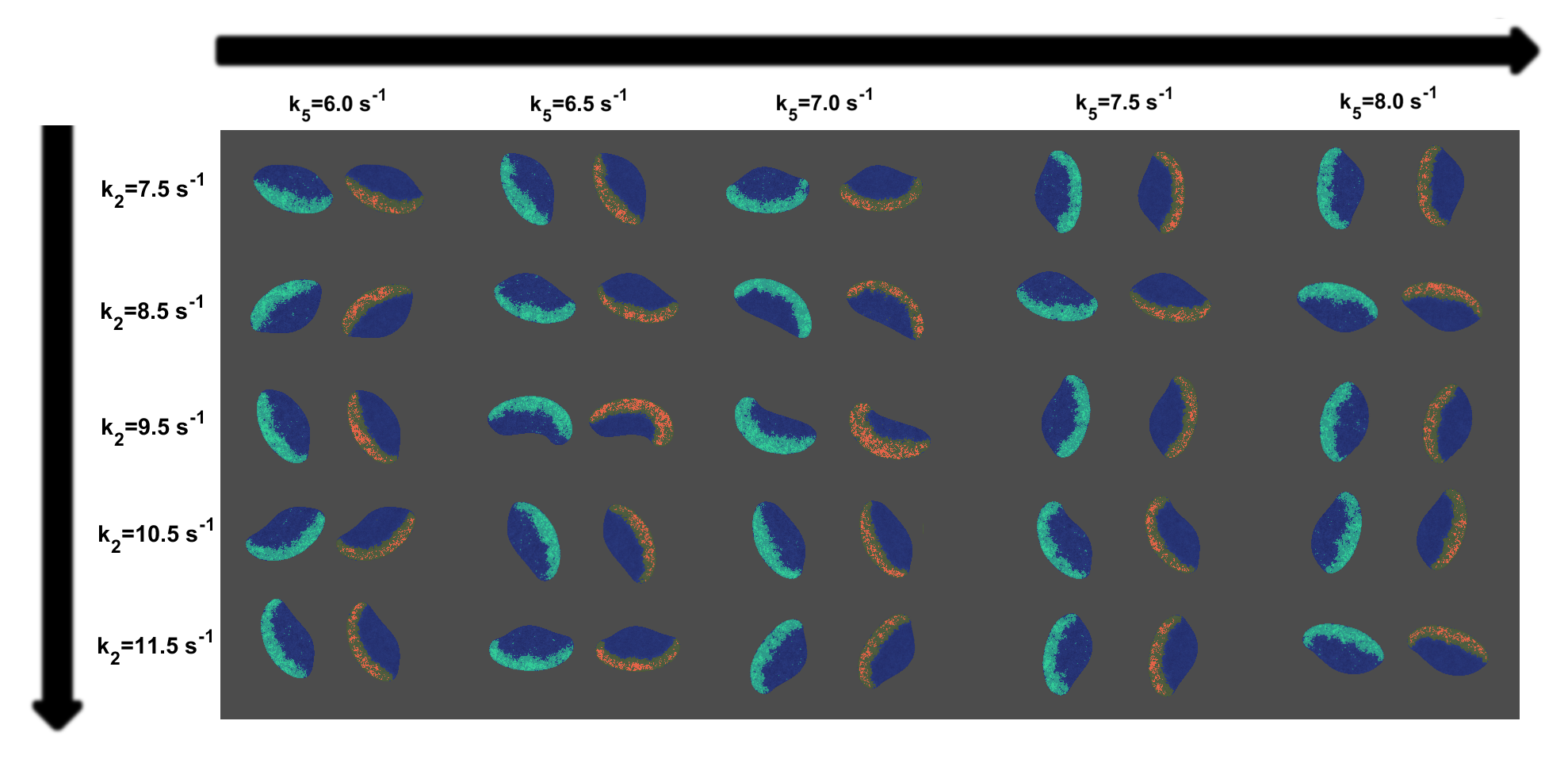}% This is a *.eps file
\end{center}
\caption{Map of snapshots of the virtual cell taken from different values of $k_2$ and $k_5$ for the new mass-conserved model. For every column, left/green snapshots refer to R distribution while right/red snapshots indicate P distribution.   
% R (in green) and P (in red) dynamics obtained from computer simulations and applying the phase field technique to the mass conserved model for different values of $k_2$ and $k_5$. 
Variances of noise intensity were fixed at $\sigma_R=0.04$ and $\sigma_P=0.025$.}
  \label{fig:fig6}
\end{figure*}

In Fig.~\ref{fig:fig6} several realizations incorporating the conservation constraint of R and P proteins, for different parameter values are shown; see Eqs.\ref{eqn:k2} and \ref{eqn:k14}. 
A direct comparison can be made with the results displayed in Fig.~\ref{fig:fig4}. 
Results in Fig.~\ref{fig:fig6} show the effects of the inclusion of the mass conservation feedback, where a wide change in the parameter values does not affect either the bistability of the model or the quantity of protein R inside the cell phase field, which was kept constant.
The inclusion of this global condition permits the use of a larger range of parameter values, increasing the robustness of the mechanism in comparison with the original model.

\begin{figure*}[t!]
\begin{center}
\includegraphics[width=\linewidth]{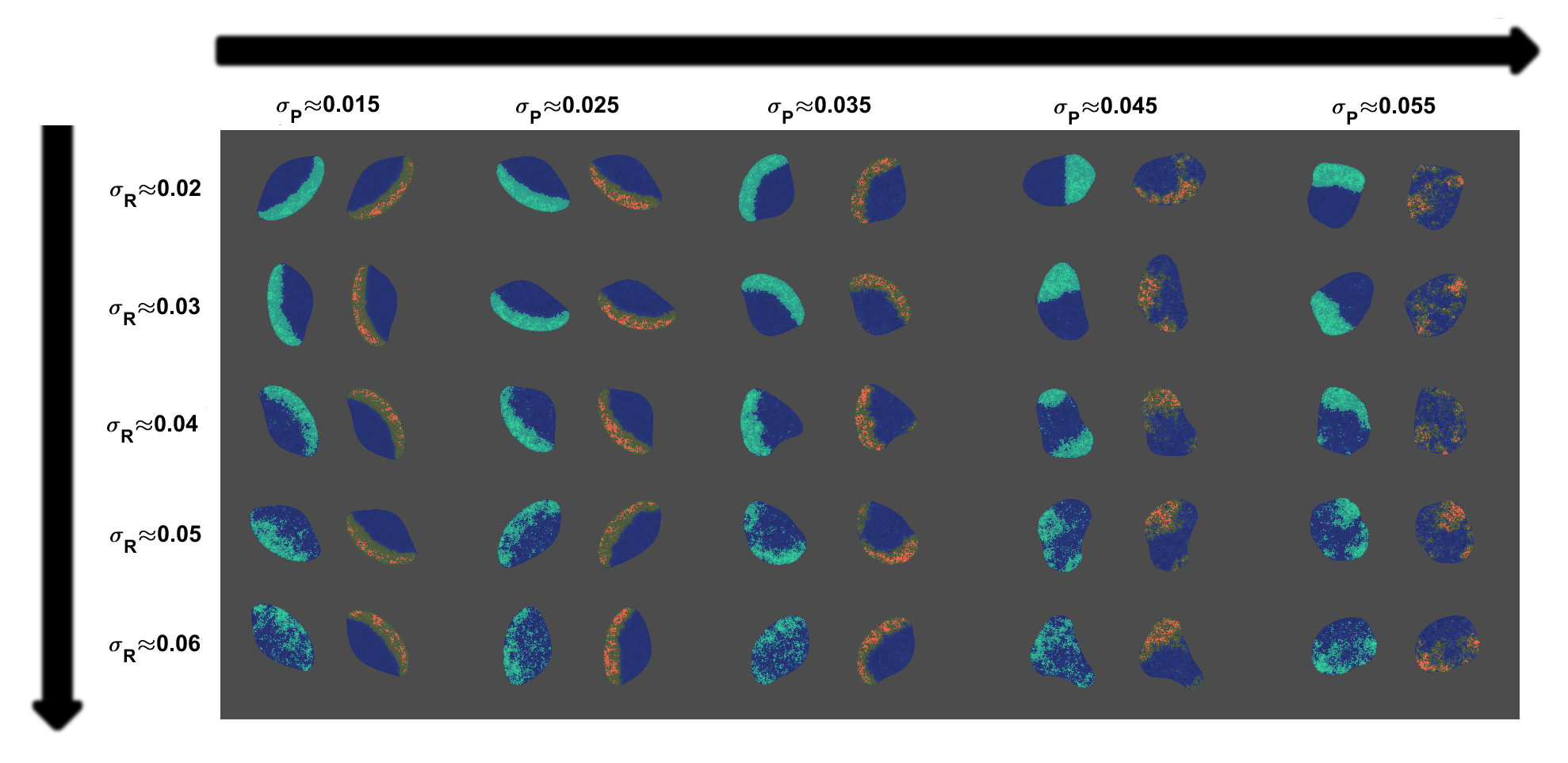}% This is a *.eps file
\end{center}
\caption{Map of snapshots of the virtual cell taken from different variance values of noise intensity for the new mass-conserved model. %Results obtained correspond to
%0.025 y 0.04 son los valores estandar del modelo de VanHaastert, solo fui variando el ruido de P y R, es decir cuando se quedo fijo P cambie R y cuando se quedo fijo R se cambio P. 
For every column, left/green snapshots refer to R distribution while right/red snapshots indicate P distribution. The value of the parameter of the simulations are $k_{2}=9.50$, $k_{5}=7.0$, $k_{14}=10.20$ and $k_{18}=7.0$. The rest of them are consistent with Table 1.}
  \label{fig:fig7}
\end{figure*}

As with the cases without mass conservation, the shape and type of motion were affected by the noise intensity. Using the same parameters of Fig.~\ref{fig:fig5} we studied the effects of varying the noise variance in the new mass-conserved model. In Fig.~\ref{fig:fig7} we see two different types of transitions. The first is observed when the variance related to the pseudopod inducer P increases, changing the shape from persistent fan-shape to random amoeboid phenotype. This change is present because while small values of P variance generates less blurred patches, an increment in the P variance drives the opposite effect, which is the appearance of more blurred and distributed patches for P.

For a second transition, we follow the same comparison line between the simulations without and with mass conservation from Fig.~\ref{fig:fig5} and \ref{fig:fig7}, respectively; now we analyze the effects for different values of the variance corresponding to the dynamics of the Ras concentration R. Similar to the patterns of P, small values of $\sigma_R$ induce the formation of less diffuse and distributed Ras patches. In contrast, greater values of $\sigma_R$ stimulate more diffuse and distributed Ras patches.
Despite the close relationship between the effects of $\sigma_R$ and $\sigma_P$, the real difference is in the dynamics of the movement. For example, if we focus on the top of the first column of Fig.~\ref{fig:fig7} for small values of the noise variances we see for both snapshots a domain in the front part of the cell which translates into persistent motion. Now, if we look in detail at the bottom of the first column corresponding to higher a value for the variance $\sigma_R$, we see a notable difference in the left hand snapshots, moving from a more compact domain in the front part of the cell to a more diffuse domain inside the cell. This transition only takes place at the Ras concentration (R) level, while maintaining consistent dynamics of the pseudopod inducer P and thus persistent motion of the cell. The same analogy could be made for the last column in Fig.~\ref{fig:fig7} where high values of $\sigma_P$ lead to a more random motion for all the cases regardless of the value of $\sigma_R$, the effect of which is to promote the appearance of less and more distributed Ras patches but without affecting the dynamics.

\subsection{Model with mass-conservation constraint is more robust}

The use of the mass conservation condition has been previously employed for the generation of stable pattern formation on the polarization process in single cells. We employed a version of the mechanism to modify previous reaction-diffusion equations to incorporate restrictions on the conservation of R and P. 
We show that this condition increases the window of parameters of the model to develop typical dynamics of cell crawling. 

\begin{figure}[h!]
\begin{center}
\includegraphics[width=\linewidth]{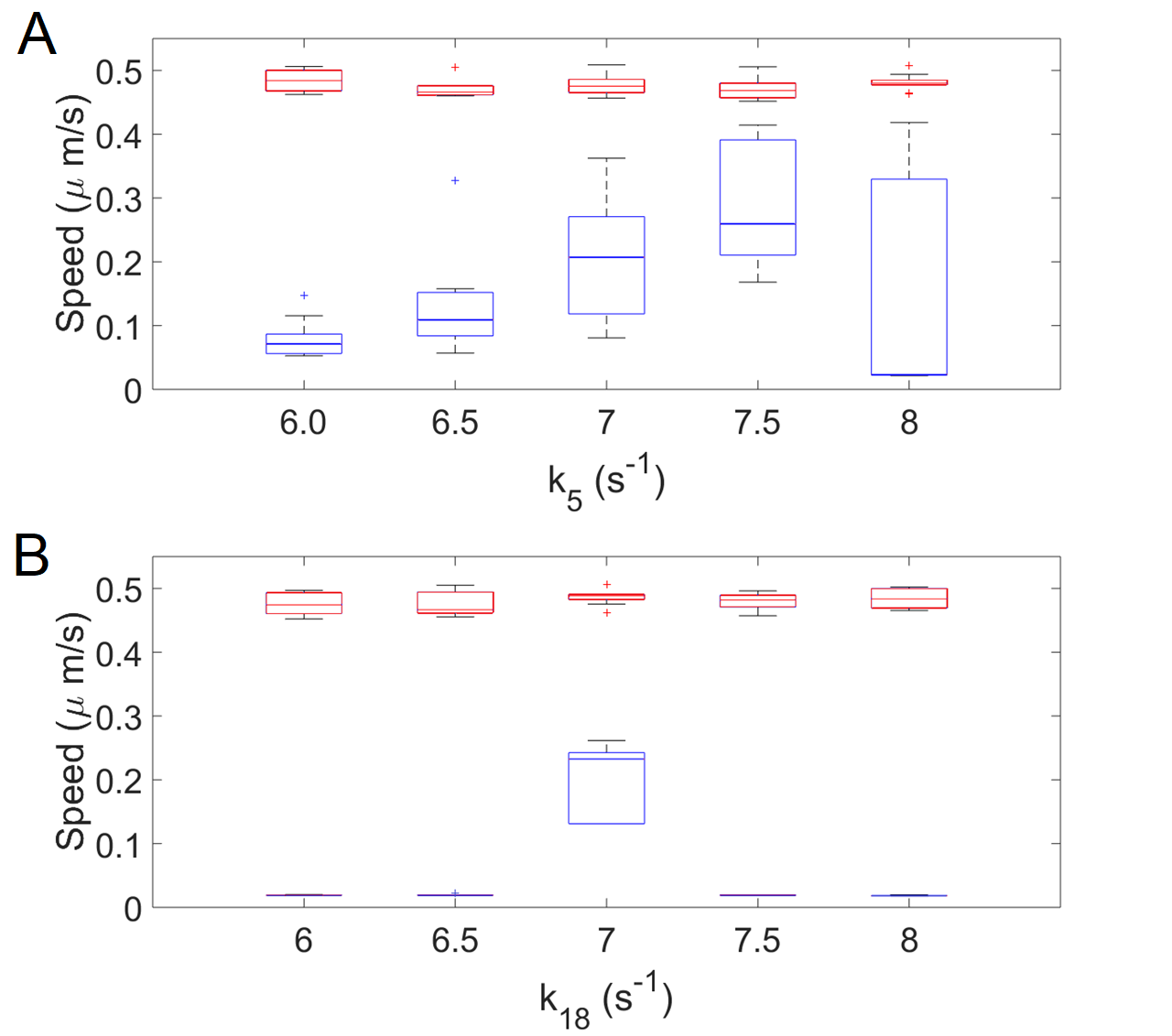}% This is a *.eps file
\end{center}
\caption{Box plot representation for the cell speed of the original model (blue color) and the new mass-conserved model (red color). Results are for when A) $k_5$ and B) $k_{18}$ are varied. For both cases it was used a fixed value of $k_2=9.5$ and $k_{14}=10.2$, $\sigma_R=0.04$ and $\sigma_P=0.025$. Ten realizations for each case were performed. }
  \label{fig:fig8}
\end{figure}

For the characterization of the mechanism we calculated the speed of the resulting cells for different parameter conditions for the original and the new mass-conserved models. 
We observed that the dynamics of the simulated cells are more robust for the mass-conserved version of the equations; see Fig.~\ref{fig:fig8}. 

%ANTES ESTABA ASI While for the non-conserved version of the model the speed and the shape drastically changes when varying the parameters and only reproduces the typical dynamics of crawling cells for a small window of parameter values

For the original model the speed and shape drastically changes when varying the parameters, and every simulation reproduces different dynamics of crawling cells (see 
%second, third and fourth snapshot of 
Fig.~\ref{fig:fig4}).
%for first, third and fifth column, respectively
%ANTESIn the other hand for the conserved version both, velocities and shapes are more realistic for a much larger window of parameter values.
Furthermore, for the new mass-conserved version both velocities and shapes are more similar to each other for a much larger window of parameter values,
%ANTES ESTABA ASI Results displayed in Fig.~\ref{fig:fig8} correspond to simulations shown in the third row of Figs.~\ref{fig:fig4} and \ref{fig:fig6}. 
as we can see in the speeds calculated in Fig.~\ref{fig:fig8}. Results displayed in Fig.~\ref{fig:fig8}A were obtained by changing $k_{5}$ to correspond to simulations shown in the third row of Figs.~\ref{fig:fig4} and \ref{fig:fig6}. The dynamics are fixed because of the conservation condition. 
In contrast, Fig.~\ref{fig:fig8}B was obtained by changing $k_{18}$ to correspond to simulations shown in the third row of Figs.~\ref{fig:fig4.5} for the no mass conservation case (results not shown for the mass-conservation system). Here one can see the preserving of the speed in the new mass-conserved system in comparison with the original model due to the small window of parameters, without the mass conservation condition that shows crawling dynamics.

%LO QUITE Y LO INCLUIR EN EL PARRAFO DE ARRIBA There is an increment in the speed from simulations without the mass conservation condition in comparison with simulations where the mass is conserved.  

%Finally, in the same way as with some parameters of the model influence in the phenotype shape. It is probably to think that a modification in the noise variance will boost certain changes in the cell velocities for both, non-conserved and conserved results. 

\begin{figure*}[t!]
\begin{center}
\includegraphics[width=\linewidth]{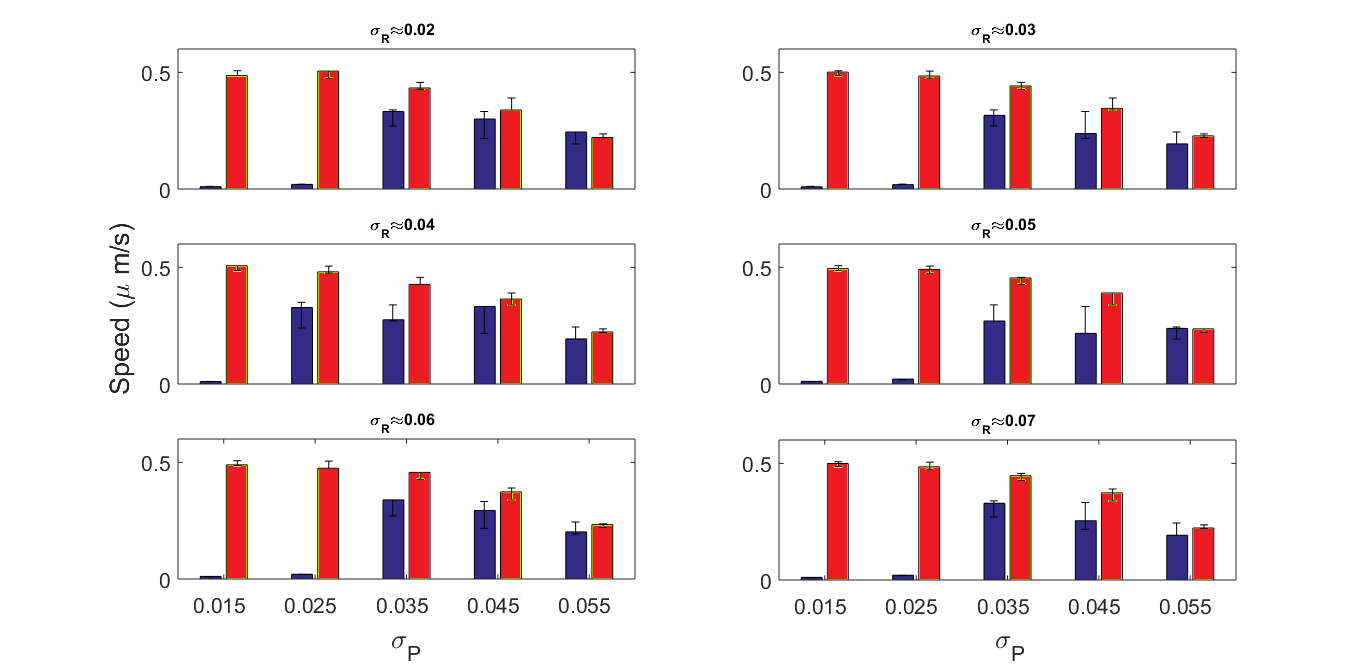}% This is a *.eps file
\end{center}
\caption{Speed bar plots measured by the original model (blue bars) and the new mass-conserved model (red bars) by varying the noise variance $\sigma_P$. In each panel the variance associated to $\sigma_R$ was fixed for different values.}
  \label{fig:fig9}
\end{figure*}

In Fig.~\ref{fig:fig9} we see the dependency of the speed on  the noise $\sigma_P$ for different values of $\sigma_R$. The speed of the cells employed for the calculation comes from the results in Figs.~\ref{fig:fig5} and \ref{fig:fig7}. For the outcome of the original model, which is displayed in blue bars, we do not observe a clear tendency in the bars. However, in contrast, in the new mass-conserved model results, represented by red bars, a clearer tendency in the bars is noted.

We observe that the speed obtained from the new mass-conserved model decreases for greater noise intensities. However, for the set of blue bars almost null velocities appear for low noise magnitude, and as we increase the noise variance, different values close to the magnitude of the red bars are reached, thereby showing a similarity in the speed measurements as the variance is increased.

%The only similarity between the two cases is the tendency for the velocity in the high noise values 

%The same trend can be observed in Figure \ref{fig:fig10} where the curves/bars are displayed by separate.

\section{Discussion}

%summary
We studied the dynamics of a one-dimensional \textit{Dictyostelium Discoideum} cell model which is mainly made up of coupled Ras activation (R), pseudopod inducer (P) and stochasticity. The deterministic model by itself reveals the emergence of bistability. This is from the visible hysteresis transition under the change of a set of parameters. In contrast, the model with the stochastic part permits the transition between two stable regimes which are induced by the parameters and the noise intensity.
The extension of the model into two dimensions with the use of the addition of a phase field, allowing cell deformation, shows similar behaviour when parameters are varied to the one-dimensional model, in agreement with \cite{van2017coupled}. 

%mass conservation
Since for most of the realistic scenarios the mass barely changes despite environmental perturbations, we chose to incorporate global feedback for the Ras and pseudopod inducer inside the cell. This increases the robustness of the model and expands the parameter range without affecting the shape and dynamics.

Note that other global feedback were already considered in the original version of the model \cite{van2017coupled}; see the dynamics of the global inhibitors, defined by G$_R$ and G$_P$ in Eq.(\ref{eqn:eqdos}) and Eq.(\ref{eqn:eqcinco}) respectively, similarly to other models of the motion of \textit{Dictyostelium discoideum} cells based on global cytosolic quantities \cite{taniguchi2013phase,xiong2010cells}. The addition of global control quantities which can determine the available concentration inside the cell has previously been employed in other models \cite{camley2017crawling,moreno2020modeling,imoto2020comparative}.

Instant speed was measured, highlighting that for the original model a considerable set of parameters driven to velocities of close to zero, while for the set of parameters for the new mass-conserved model the speed remained almost constant. 
%Even though the main change is observed when different noise intensities were used in the model. Manifesting more magnitudes different from zero for the no conserved case and a clear tendency to achieve small magnitudes of speed as we increase the variance of the pseudopod inducer $\sigma_P$

Note that we have shown that the original model is based on a bistable conditions with the appropriate noise intensity. We have previously used simple stochastic bistable models for the description of the motion of \textit{Dictyostelium discoideum} cells \cite{alonso2018modeling,moreno2020modeling} based on the same concept together with a mass-conservation constraint. There are also other models based on similar constraints \cite{imoto2020comparative}. Therefore, this constraint is a convenient mechanism to increase robustness in the delicate equilibrium between stochastic fluctuations and a bistable condition.

\begin{figure}[h!]
\begin{center}
\includegraphics[width=\linewidth]{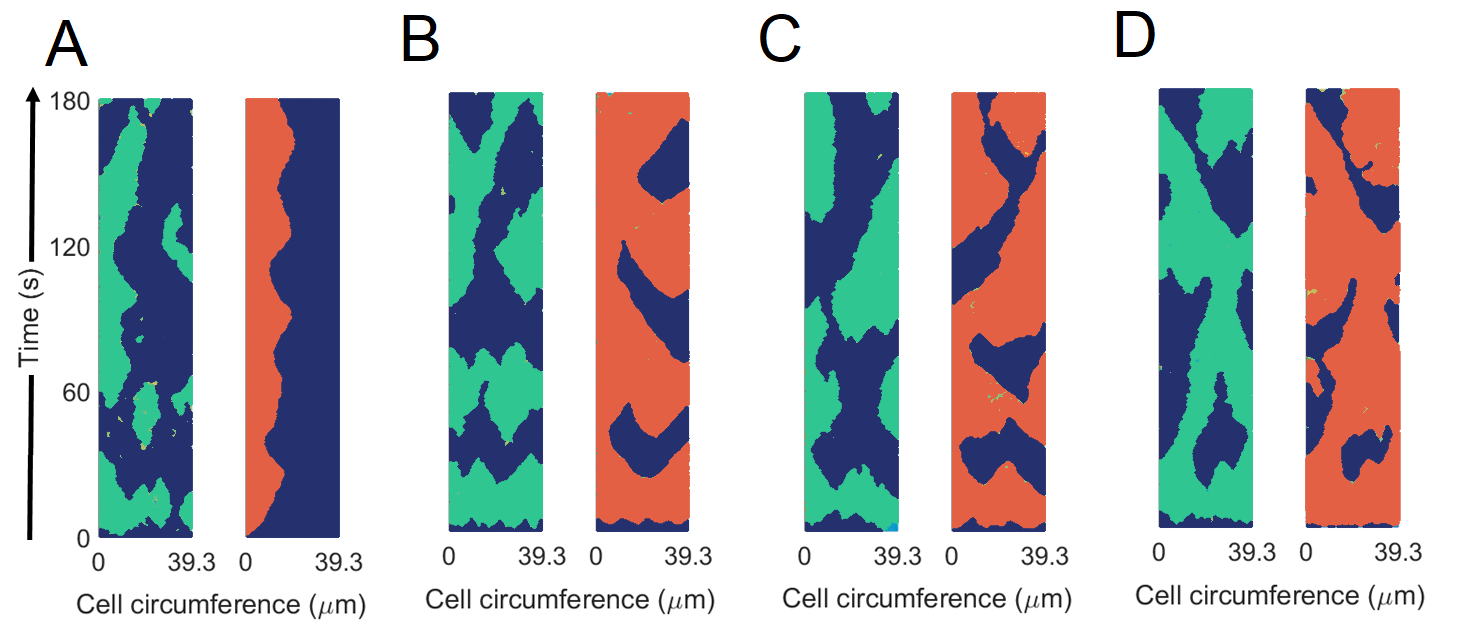}% This is a *.eps file
\end{center}
\caption{Spatio-temporal plots of R (in green) and P (in orange) for different values of the coupling parameter between R and P: $k_{13} = 0.1$ (A), $k_{13} = 0.50$ (B), $k_{13} = 1$ (C), and $k_{13} = 2$ (D) in the new mass-conserved model. Results are for a fixed values  of $\sigma_R=0.04$ and $\sigma_P=0.025$. }
  \label{fig:fig10}
\end{figure}

%NEW PARAGRAPH
The coupling of R on the pseudopod inducer P is weak. 
The coupling is produced by the term proportional to $k_{13}$ and it increases the probability of the generation of patches of P. A chemotactic concentration can enhance R and, therefore, polarize the field P and direct the locomotion \cite{van2017coupled}. In the absence of chemotaxis the coupling is small, while more physiological conditions may require larger couplings. In Fig.\ref{fig:fig10} we increased parameter $k_{13}$ to demonstrate the increase in the coupling of both field R and P through the parameter. Intermediate values can be compared with experimental measures to fit an appropriate value; however such evaluation is outside the scope of this work.
%NEW PARAGRAPH END

An extension of the new mass-conserved model is possible with the inclusion of an external chemo-attractant gradient to analyze the response, known as chemotaxis and already considered in the original model \cite{van2017coupled}. Future simulations should study the relation of the chemotactic motion of \textit{Dictyostelium discoideum} cells  with the stochastic fluctuations and the rest of the parameters. They should also measure the effects on the shape and speed caused by the gradient.

%final statement
In summary, we have shown that the inclusion of a mass-conservation constraint in the correct position substantially increases the robustness of a computational model of a crawling cell.

\section*{Acknowledgments}

We thank grant PGC2018-095456-B-I00 funded by MCIN/AEI/10.13039/501100011033 and by “ERDF A way of making Europe”, by the “European Union”.
E.M. acknowledges also financial support from CONACYT.

\appendix
\section{Integration of stochastic partial differential equations}
\label{appendixA}

The Gaussian white noises employed in the model described here are always additive, and therefore it is enough to use a Euler-Maruyama method for the integration of the stochastic partial differential equations (SPDE) considered here\cite{garcia2012noise}. Assuming a generic version of the equations considered here:
\begin{eqnarray}
\label{eqn:equnoap}
\frac{\partial C(\vec{x},t)}{\partial t}=  F(C(\vec{x},t)) 
+ \xi(\vec{x},t),
\end{eqnarray}
the standard method to integrate this SPDE consists in the discretization in space ($\Delta \vec{x}$): from  $C(\vec{x},t)$ to $C_i(t)$, and time ($\Delta t$) and to account for the particularities of the noise terms: 
\begin{eqnarray}
C_i(t+\Delta t) = C_i(t) + \Delta t F(C_i(t)) + \sqrt{\frac{2 \sigma^2 \Delta t}{\Delta x^d} } 
N(0,1), \nonumber \\
\label{eqn:eqdosap}
\end{eqnarray}
where d is the spatial dimensionality of the domain considered (here d=1 or d=2), and N(0,1) corresponds to a random number generated by a normal distribution of mean zero and variance 1. 

Such standard method for the integration of SPDE is equivalent under some conditions with the version employed in the original publication of the model \cite{van2017coupled} employed here:
\begin{eqnarray}
C_i(t+\Delta t) = C_i(t) + \Delta t F(C_i(t)) + 
N(0,\sigma_2), \nonumber \\
\label{eqn:eqtresap}
\end{eqnarray}
which under correct tuning of the parameter $\sigma_2$ can give rise to an equivalent description.

%Puntos Pendientes?
%mov rotacional en circulo 
%cambiar anti alignment alignment
%pushing cambiarlo a dos tipos 
%error bars
%Number of realizations captions figure 7
%Poner los segundos de ralizaciones
 
%\newpage
 
\section{Parameters and constants used in the model.}
The meaning of the biochemical parameters and of the parameters associated to the phase field can be obtained, respectively from the original studies \cite{van2017coupled} and \cite{moreno2020modeling}.

Note also that the parameters related to a chemoattractant gradient \cite{van2017coupled} were excluded in this study.

\begin{table}[h!]
\begin{tabular}{ |c|c||c|c| } \hline
Parameter & Value & Parameter & Value \\
 \hline
 $k_{0}$ &  0.01 $s^{-1}$ &  $D_{R}$ &  0.75 $\mu m^2/s$\\
 $k_{1}$ & 1.5 $s^{-1}$ & $D_{LR}$ &  0.25 $\mu m^2/s$\\
 $k_{2}-k_{2}^*$ & 7.5-11.5 $s^{-1}$ &  $D_{P}$ &  0.75 $\mu m^2/s$\\ 
 $k_{3}$ & 1.5 $s^{-1}$ & $D_{LP}$ &  0.25 $\mu m^2/s$\\ 
 $k_{4}$ & 1.0 $s^{-1}$ & $D_{M}$ &  0.25 $\mu m^2/s$\\
 $k_{5}$ & 6.0-8.0 $s^{-1}$ & $n_1$ & 2.0\\
 $k_{6}$ & 0.1 $s^{-1}$ & $n_2$ & 2.0\\
 $k_{7}$ & 0.05 $s^{-1}$ & $n_3$ & 2.0\\
 $k_{9}$ & 0.15 $s^{-1}$ & $K_R$ & 0.4\\
 $k_{10}$ & 0.1 $s^{-1}$ & $K_P$ & 0.4\\
 $k_{11}$ & 0.1 $s^{-1}$ & $K_M$ & 0.4\\
 $k_{12}$ & 0.1 $s^{-1}$ & $\sigma_R$ &  $0.02-0.06$ \\
 $k_{13}$ & 0.1 $s^{-1}$ & $\sigma_P$ &  $0.015-0.055$\\
 $k_{14}-k_{14}^*$ & 9.2-11.2 $s^{-1}$ & $\tau$ & 2.0 $pNs\mu m^{-2}$\\
 $k_{15}$ & 0.5 $s^{-1}$  & $\gamma$ & 2.0 $pN$ \\
 $k_{16}$ & 1.5 $s^{-1}$  & $\epsilon$ & 0.75 $\mu m$  \\
 $k_{17}$ & 1.8 $s^{-1}$  & $\beta$ & 18.64 $pN \mu m^{-3}$  \\
 $k_{18}$ & 6.0-8.0 $s^{-1}$  & $A_0$ &  122.7 $\mu m^2$  \\
$k_{19}$ & 0.1 $s^{-1}$  & $\alpha$  & 10.0 $pN\mu m^{-1}$  \\
$k_{20}$ & 0.15 $s^{-1}$   & $C_R$ &  30.6 $\mu m^2$  \\
$k_{21}$ & 0.2 $s^{-1}$   & $C_P$ &  30.6 $\mu m^2$  \\
$k_{22}$ & 0.1 $s^{-1}$   & $\eta$ &  $2.0 s^{-1}$  \\
$k_{23}$ & 0.1 $s^{-1}$   & $\Delta t$ &  $0.03-0.003$ $s$  \\
$k_{24}$ & 0.03 $s^{-1}$  & $\Delta x$ &  $0.32-0.16$ $\mu m$  \\
 \hline
\end{tabular}
\label{table:1}
\end{table}

%\begin{thebibliography}{4}
\newpage
\bibliography{AmoeboidCells}
%\end{thebibliography}

\end{document}